\documentclass[aps,pre,groupedaddress,twocolumn]{revtex4}

\usepackage{graphicx}

\begin{document}

\renewcommand{\vec}[1]{{\mathbf #1}}

\title{Langevin description of speckle dynamics\\ in nonlinear disordered media}
\author{S.E. Skipetrov}
\email[]{Sergey.Skipetrov@grenoble.cnrs.fr}
\affiliation{Laboratoire de Physique et Mod\'elisation des Milieux Condens\'es,\\
CNRS, 38042 Grenoble, France}

\date{\today}

\begin{abstract}
We formulate a Langevin description of dynamics of a speckle pattern
resulting from the multiple scattering of a coherent wave in a nonlinear
disordered medium.
The speckle pattern exhibits instability with respect to
periodic excitations at frequencies $\Omega$ below some $\Omega_{\mathrm{max}}$,
provided that the nonlinearity exceeds some $\Omega$-dependent threshold.
A transition of the speckle pattern from a
stationary state to the chaotic evolution is predicted
upon increasing nonlinearity.
The shortest typical time scale of chaotic intensity fluctuations is of
the order of $1/\Omega_\mathrm {max}$.
\end{abstract}


\maketitle


Propagation of a coherent wave in a disordered medium is diffusive if $\lambda \ll l \ll L$, where
$\lambda$ is the wavelength, $l$ is the mean free path, and $L$ is the size of the medium \cite{rossum99}.
While the wave undergoes multiple scattering and the spatial distribution of the scattered
intensity looks quite irregular (speckle pattern), the coherence of the wave is not destroyed and various
coherent phenomena can be observed: enhanced backscattering, short- and long-range intensity correlations,
universal conductance fluctuations, etc. (see Refs.\ \onlinecite{rossum99,berk94,sebbah01} for reviews).
Available studies of {\em nonlinear} phenomena for diffuse waves include calculations of the
enhanced backscattering cone at fundamental \cite{agran91} and doubled \cite{kravtsov91} frequencies,
investigations of optical phase conjugation \cite{kravtsov90}, studies of correlations in transmission
and reflection coefficients of the second harmonic \cite{boer93} and fundamental \cite{bress00} waves,
an extension of the standard diagrammatic technique to nonlinear disordered media
\cite{wonderen94}, and a study of persistent hole burning in multiple-scattering
media \cite{tomita01}.

After realizing that the sensitivity of the speckle pattern to changes of the scattering potential diverges for a
sufficiently strong nonlinearity \cite{spivak00}, a new phenomenon, the temporal instability of the
multiple-scattering speckle pattern in a disordered medium with cubic nonlinearity, has recently
been predicted \cite{skip00}.
The speckle pattern is expected to become unstable and to exhibit spontaneous fluctuations if the nonlinearity
exceeds some critical value.
Although of primary importance in view of the possible experimental observation of the instability
phenomenon, the dynamics of spontaneous intensity fluctuations, their nature and associated
characteristic time scales have not yet been studied up to now.

In the present paper we formulate the \textit{dynamic} Langevin description of
spontaneous intensity fluctuations in a nonlinear disordered medium.
Our theoretical method can be viewed as an extension of the \textit{stationary}
Langevin approach introduced in Ref.\ \onlinecite{spivak00}, the latter being inadequate
to describe the dynamics of speckles.
Analysis of the speckle pattern stability
with respect to weak periodic excitations shows that if the effective nonlinearity
parameter $p = \Delta n^2 (L/l)^3$ exceeds some critical value $p_\mathrm{c} \simeq 1$
(where $\Delta n$ is the typical value of the nonlinear correction to the refractive index) the speckle pattern
becomes unstable with respect to periodic excitations at frequencies inside some limited
low-frequency interval, and the maximal Lyapunov exponent becomes positive.
This allows us to describe the chaotic nature of spontaneous intensity fluctuations beyond
the absolute instability threshold $p = p_\mathrm{c}$ and to estimate their characteristic time scale.

We consider a scalar wave propagating in a nonlinear disordered medium and described
by the following wave equation:
\begin{eqnarray}
&&\left\{ \nabla^2 - \frac{1}{c^2} \frac{\partial^2}{\partial t^2}
\left[1 + \varepsilon(\vec{r}, t) + \varepsilon_2 \left| \psi(\vec{r}, t)\right|^2 \right]
\right\} \psi(\vec{r}, t)
\nonumber \\
&& = J(\vec{r}, t),
\label{weq}
\end{eqnarray}
where
$J(\vec{r}, t) = J_0(\vec{r}) \exp(-i \omega_0 t)$ is a monochromatic source term,
$c$ denotes the speed of wave in the average medium,
$\varepsilon(\vec{r}, t)$ is the fractional
fluctuation of the dielectric constant at frequency $\omega_0$
(possibly slowly varying in time), and $\varepsilon_2$ is a nonlinear constant.
Equation (\ref{weq}) describes, e.g., propagation of optical waves in media
with intensity-dependent refractive index \cite{shen84} in the scalar approximation
and neglecting the generation of the third optical harmonics. The latter assumption
is justified in the absence of phase matching \cite{shen84}
or, more precisely, when $\left| k(3 \omega_0) - 3 k(\omega_0) \right| l \gg 1$, where
$k(\omega)$ is the wavenumber at frequency $\omega$.

\begin{figure}
\includegraphics[width=7cm]{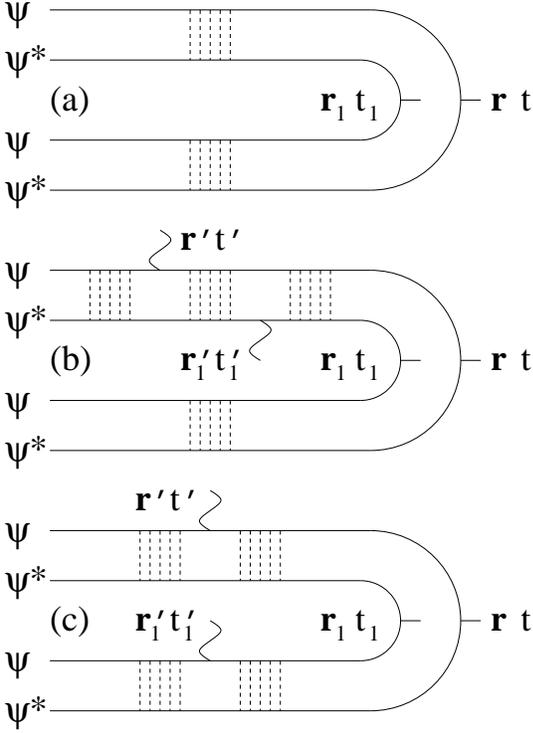}
\caption{\label{fig1} Diagrams contributing to Eqs.\ (\ref{current}) and (\ref{qcorr}).
Solid lines denote the wave field $\psi$ and the complex conjugated field $\psi^*$.
Dashed lines denote scattering of $\psi$ and $\psi^*$ on the same heterogeneity.
The diagrams (b) and (c) are obtained by inserting two $k_0^2$ vertices
(denoted by wavy lines) to the diagram (a) at ($\vec{r}^{\, \prime}, t^{\, \prime}$) and
($\vec{r}_1^{\, \prime}, t_1^{\, \prime}$), respectively.}
\end{figure}

Consider first a linear medium ($\varepsilon_2 = 0$) of typical size $L$ and a white-noise Gaussian disorder:
$\left< \varepsilon(\vec{r}, t) \varepsilon(\vec{r}_1, t) \right> =
4 \pi / (k_0^4 l) \delta( \vec{r} - \vec{r}_1)$, where
$k_0 = k(\omega_0) = \omega_0 / c$.
Let the time variations of
$\varepsilon(\vec{r}, t)$ be random, stationary, and arbitrary slow,
so that the time scale of the resulting variations of
the amplitude $\varphi(\vec{r}, t)$ of
$\psi(\vec{r}, t) = \varphi(\vec{r}, t) \exp(-i \omega_0 t)$ is much larger than the typical time between
two successive scattering events $l/c$.
For $L \gg l$ and far enough from the boundaries
of the disordered sample,
the average intensity $\left< I(\vec{r}) \right>$
then obeys the diffusion equation \cite{ishim81}, while the long-range correlation of intensity fluctuations
$\delta I(\vec{r}, t) = I(\vec{r}, t) - \left< I(\vec{r}) \right>$ can be found by solving the
Langevin equation \cite{zyuzin87}:
\begin{eqnarray}
\frac{\partial}{\partial t} \delta I(\vec{r}, t)  - D \nabla^2  \delta I(\vec{r}, t) =
- \vec{\nabla} \cdot {\vec j}_\mathrm{ext}(\vec{r}, t),
\label{langevin}
\end{eqnarray}
where $I(\vec{r}, t) = \left| \varphi(\vec{r}, t) \right|^2$, $D = c l/3$ is the diffusion constant,
and ${\vec j}_\mathrm{ext}(\vec{r}, t)$ are random external Langevin currents:
\begin{eqnarray}
&&\left< j_\mathrm{ext}^{(i)} (\vec{r}, t) j_\mathrm{ext}^{(j)} (\vec{r}_1, t_1) \right>
\nonumber \\
&&=
2 \pi l c^2 / (3 k_0^2) \left| \left< \varphi(\vec{r}, t) \varphi^*(\vec{r}, t_1) \right> \right|^2
\delta_{ij} \delta(\vec{r} - \vec{r}_1).
\label{current}
\end{eqnarray}
The diagram corresponding to Eq.\ (\ref{current}) is shown in Fig\ \ref{fig1}(a).

For a given $J_0(\vec{r})$, the current
$\vec{j}_\mathrm{ext} (\vec{r}, t)$ is a ``fingerprint'' of the disorder $\varepsilon(\vec{r}, t)$.
An infinitesimal variation $\Delta \varepsilon(\vec{r}, t)$ of the dielectric constant will modify
$\vec{j}_\mathrm{ext} (\vec{r}, t)$  by a small amount
\begin{eqnarray}
&&\Delta \vec{j}_\mathrm{ext} (\vec{r}, t)
\nonumber \\
&&= \int_V \mathrm{d}^3 \vec{r}^{\, \prime} \int_{-\infty}^{t}
\mathrm{d} t^{\, \prime} \,
\vec{q}( \vec{r}, \vec{r}^{\, \prime}, t - t^{\, \prime} ) \; \Delta \varepsilon(\vec{r}^{\, \prime}, t^{\, \prime}),
\label{dj}
\end{eqnarray}
where the spatial integral is over the volume $V$ of the sample, we neglect
the terms of the second and higher orders in $\Delta \varepsilon(\vec{r}, t)$,
and the correlation of
random response functions $\vec{q}( \vec{r}, \vec{r}^{\, \prime}, \Delta t = t - t^{\, \prime}) =
\delta \vec{j}_\mathrm{ext}(\vec{r}, t) / \delta \varepsilon(\vec{r}^{\, \prime}, t^{\, \prime})$
can be found by a functional differentiation of Eq.\ (\ref{current}):
\begin{eqnarray}
&&\left< q^{(i)}( \vec{r}, \vec{r}^{\, \prime}, \Delta t)
q^{(j)}( \vec{r}_1, \vec{r}_1^{\, \prime}, \Delta t_1) \right>
\nonumber \\
&&=
3 \pi D^2 (c^2/l) \delta_{ij} \delta(\vec{r} - \vec{r}_1)
\nonumber \\
&&\times \left[
\left< I(\vec{r}^{\, \prime}) \right> G(\vec{r}^{\, \prime}, \vec{r}_1^{\, \prime}; \Delta t -\Delta t_1)
G(\vec{r}_1^{\, \prime}, \vec{r}; \Delta t_1) \left< I(\vec{r}) \right> \right.
\nonumber \\
&&+ \left. \left< I(\vec{r}_1^{\, \prime}) \right> G(\vec{r}_1^{\, \prime}, \vec{r}^{\, \prime}; \Delta t_1 - \Delta t)
G(\vec{r}^{\, \prime}, \vec{r}; \Delta t) \left< I(\vec{r}) \right> \right.
\nonumber \\
&&- \left. \left< I(\vec{r}^{\, \prime}) \right> G(\vec{r}^{\, \prime}, \vec{r}; \Delta t)
\left< I(\vec{r}_1^{\, \prime}) \right> G(\vec{r}_1^{\, \prime}, \vec{r}; \Delta t_1) \right].
\label{qcorr}
\end{eqnarray}
Here $\left| \vec{r} - \vec{r}^{\, \prime} \right|$, $\left| \vec{r} - \vec{r}_1^{\, \prime} \right|$,
$\left| \vec{r}^{\, \prime} - \vec{r}_1^{\, \prime} \right| \gg l$ is assumed and
$G(\vec{r}, \vec{r}_1; \Delta t)$ is the Green's function of Eq.\ (\ref{langevin}).
The diagrams contributing to Eq.\ (\ref{qcorr}) are shown in Fig.\ \ref{fig1}(b, c).
In the stationary limit
$\Delta \varepsilon(\vec{r}, t) \equiv \Delta \varepsilon(\vec{r})$,  Eqs.\ (\ref{dj}) and (\ref{qcorr})
reduce to Eqs.\ (6) and (7) of Ref.\ \onlinecite{spivak00}.

The Langevin description of intensity fluctuations in disordered media can be extended to the nonlinear
case ($\varepsilon_2 \ne 0$). To this end, we consider time-independent $\varepsilon$:
$\varepsilon(\vec{r}, t) = \varepsilon(\vec{r})$, and assume that the diffusion
constant $D$ and the mean free path $l$ are not affected by the nonlinearity. The latter assumption is
valid if $\Delta n^2 k_0 l \ll 1$ \cite{spivak00},
where $\Delta n = (\varepsilon_2/2) I_0$ is the typical value of the nonlinear
correction to the refractive index and $I_0 \simeq \left< I( \vec{r} ) \right>$ is the typical value of the average
intensity in the medium.
We now admit that in a nonlinear medium, the total dielectric constant
contains a \textit{linear} contribution $1 + \varepsilon(\vec{r})$ that we assumed
to be time-independent, and a \textit{nonlinear}
contribution $\varepsilon_2 I(\vec{r}, t)$ that can vary with time.
The variation of the total dielectric constant can be therefore only due its
nonlinear part, and we can identify
the infinitesimal variation of the dielectric constant
$\Delta \varepsilon(\vec{r}^{\, \prime}, t^{\, \prime})$ in Eq.\ (\ref{dj})
with $\varepsilon_2 \Delta I(\vec{r}^{\, \prime}, t^{\, \prime})$,
where $\Delta I(\vec{r}^{\, \prime}, t^{\, \prime})$ is the change of the intensity
at $\vec{r}^{\, \prime}$ during some infinitesimal time interval
$(t^{\, \prime}, t^{\, \prime} + \delta t)$:
$\Delta I(\vec{r}^{\, \prime}, t^{\, \prime}) =
I(\vec{r}^{\, \prime}, t^{\, \prime} + \delta t) -
I(\vec{r}^{\, \prime}, t^{\, \prime})$.
Substituting
$\Delta \varepsilon(\vec{r}^{\, \prime}, t^{\, \prime}) =
\varepsilon_2 \Delta I(\vec{r}^{\, \prime}, t^{\, \prime})$ into Eq.\ (\ref{dj}),
noting that
$\Delta \vec{j}_\mathrm{ext} (\vec{r}, t) =
\vec{j}_\mathrm{ext} (\vec{r}, t+\delta t) - \vec{j}_\mathrm{ext} (\vec{r}, t)$,
and dividing both sides of the resulting equation by
$\delta t \rightarrow 0$, we obtain the following dynamic equation:
\begin{eqnarray}
\frac{\partial}{\partial t} \vec{j}_\mathrm{ext} (\vec{r}, t) &=& \varepsilon_2
\int_V \mathrm{d}^3 \vec{r}^{\, \prime} \int_{0}^{\infty} \mathrm{d} \Delta t \,
\vec{q}( \vec{r}, \vec{r}^{\, \prime}, \Delta t) \;
\nonumber \\
&\times& \frac{\partial}{\partial t} \delta I(\vec{r}^{\, \prime}, t - \Delta t),
\label{djnl}
\end{eqnarray}
where we make use of the fact that
$I(\vec{r}, t) = \left< I(\vec{r}) \right> + \delta I(\vec{r}, t)$
and hence
$\Delta I(\vec{r}, t) = \Delta [\delta I(\vec{r}, t)]$, since
$\left< I(\vec{r}) \right>$ is time-independent.

Equations (\ref{langevin}) and (\ref{djnl}) form
a self-consistent
system of equations: Eq.\ (\ref{langevin}) governs the spatio-temporal evolution
of the intensity fluctuations $\delta I(\vec{r}, t)$ due to the Langevin
currents ${\vec j}_\mathrm{ext}(\vec{r}, t)$, while Eq.\ (\ref{djnl}) describes
the distributed feedback mechanism, leading to variations of
${\vec j}_\mathrm{ext}(\vec{r}, t)$ depending on the changes of $\delta I(\vec{r}, t)$.
Note that Eq.\ (\ref{djnl}) is a linearized equation:
only the terms linear in the nonlinear contribution to the dielectric
constant $\varepsilon_2 I(\vec{r}, t)$ are kept, which is
justified as long as $\varepsilon_2 I(\vec{r}, t) \ll 1$.
In certain circumstances (see below),
the linearized nature of Eq.\ (\ref{djnl}) may result in the exponential growth of
its solution with time, and in this sense
Eqs.\ (\ref{langevin}) and (\ref{djnl}) are analogous to the equations of
linear stability analysis commonly used to study the stability of nonlinear systems
(see, e.g., Refs.\  \onlinecite{gibbs85} and
\onlinecite{arecchi99} for examples of nonlinear optical systems exhibiting
instabilities). Hence, although Eqs.\ (\ref{langevin}) and
(\ref{djnl}) allow us to study the stability of the speckle pattern and
the characteristic time scales of spontaneous intensity fluctuations beyond the
instability threshold,
they cannot be used to determine the amplitude of these fluctuations.

\begin{figure}
\includegraphics[width=8cm]{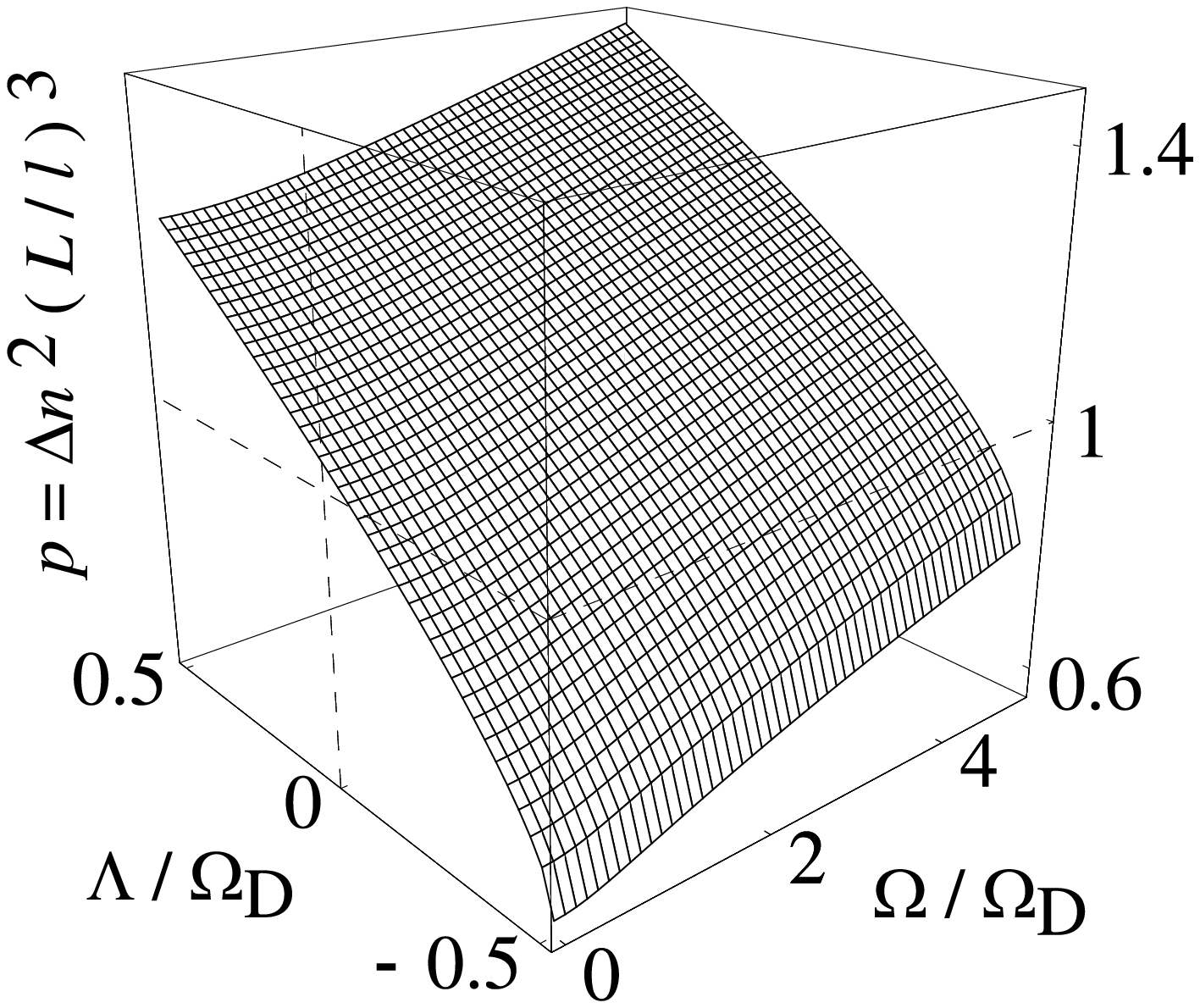}
\caption{\label{fig2} Surface describing the stability of the multiple-scattering speckle pattern in a nonlinear
disordered medium with open boundaries.
At given effective nonlinearity parameter $p$ and frequency $\Omega$, the surface
defines the Lyapunov exponent $\Lambda$. If $\Lambda > 0$, the speckle pattern is
unstable with respect to periodic excitations at frequency $\Omega$.}
\end{figure}

Consider now an infinitesimal periodic excitation of the static speckle pattern:
$\delta I (\vec{r}, t) =  \delta I (\vec{r}, \nu) \exp( i \nu t )$,
where $\nu = \Omega - i \Lambda \ne 0$ and $\Omega > 0$.
Such an excitation can be either damped or
amplified, depending on the sign of the Lyapunov exponent $\Lambda$.
The value of $\Lambda$ is determined by two competing processes: on the one hand,
diffusion tends to smear the excitation out, while on the other hand, the
distributed feedback sustains its existence.
The mathematical description of this competition is provided by
Eqs.\ (\ref{langevin}) and (\ref{djnl}), that after the substitution of
$\delta I (\vec{r}, t) =  \delta I (\vec{r}, \nu) \exp( i \nu t )$
[and similarly for $\vec{j}_\mathrm{ext} (\vec{r}, t)$] lead
to the following equation (see Appendix \ref{appa} for the details of derivations):
\begin{eqnarray}
p \simeq F \left( \Omega/\Omega_D, \Lambda/\Omega_D \right).
\label{final}
\end{eqnarray}
Here $p = \Delta n^2 (L/l)^3$ is the effective nonlinearity parameter,
the function $F$ is shown in Fig.\ \ref{fig2}, and a numerical factor
of order unity is omitted. To obtain Eq.\ (\ref{final}),
we have assumed the disordered sample
to have open boundaries (i.e. the diffusing wave leaves the sample when it reaches a boundary)
and have taken the limits of  large sample size ($L/l \gg k_0 l$) and moderate
frequency $\Omega \ll \Omega_D [L / (k_0 l^2)]^2$, where $\Omega_D = D/L^2$ is
the inverse of the typical time needed for a multiple-scattered wave to diffuse
through the disordered sample.

It follows from Fig.\ \ref{fig2} that
for a given frequency $\Omega$ the sign of the Lyapunov exponent $\Lambda$ depends on the value of $p$.
Excitations at frequencies $\Omega$ corresponding to $\Lambda < 0$ are damped exponentially and thus
soon disappear. In contrast, excitations at frequencies $\Omega$ corresponding to $\Lambda > 0$
are exponentially amplified, which signifies the instability of
the speckle pattern with respect to excitations at such frequencies.
Noting that $\Lambda$  is always negative for $p < 1$, we conclude that
all excitation are damped in this case and the
speckle pattern is absolutely stable. In an experiment, any spontaneous excitation of the static speckle
pattern will be suppressed and the speckle pattern will be independent of time:
$\delta I(\vec{r}, t) = \delta I(\vec{r})$, as in the linear case.
When $p > 1$, an interval of frequencies $0 < \Omega <
\Omega_\mathrm{max}$ starts to open up with $\Lambda > 0$.
The speckle pattern thus becomes unstable with respect to
excitations at low frequencies. In an experiment,
any spontaneous excitation of the static
speckle pattern at frequency $\Omega \in (0, \Omega_\mathrm{max})$
will be amplified and one will observe a time-varying speckle pattern $\delta I(\vec{r}, t)$.

\begin{figure}
\includegraphics[width=8cm]{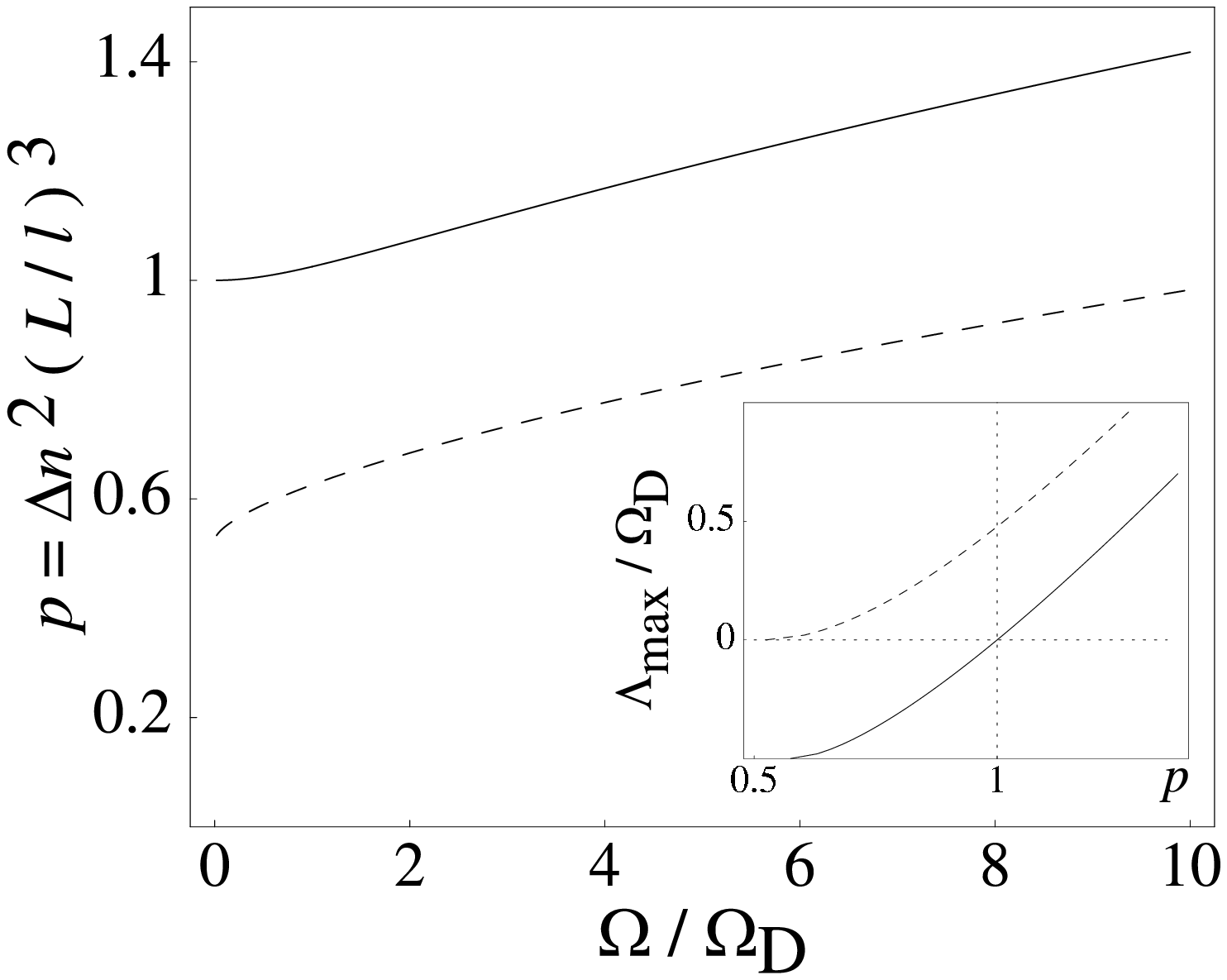}
\caption{\label{fig3} Main plot: frequency-dependent ``phase diagram'' of the multiple-scattering speckle pattern in
a nonlinear disordered medium with open (solid line) or reflecting (dashed line) boundaries.
For a given $\Omega$, $p$ should exceed the plotted threshold value
for the instability to develop.
Inset: maximal Lyapunov exponent as a function of the effective nonlinearity
parameter $p$.
The dotted lines show $\Lambda_{\mathrm{max}} = 0$ and $p = 1$.}
\end{figure}

The border between stable ($\Lambda < 0$) and unstable ($\Lambda > 0$)
regimes is shown in Fig.\ \ref{fig3} by a solid line. The instability threshold increases with $\Omega$.
At $\Omega \ll \Omega_D$ the exact functional dependence of the threshold on
$\Omega$ is rather sensitive to the peculiarities of the disordered sample (e.g., its geometry and conditions
on the boundaries), since for such slow oscillations the feedback mechanism is ensured by partial
waves that have long path lengths $s \agt L^2/l$ and hence ``feel'' the presence of the boundaries and the
shape of the sample. Using the analytic expression for the function
$F \left( \Omega/\Omega_D, \Lambda/\Omega_D \right)$
derived in Appendix \ref{appa}, we find
$p - 1 \sim (\Omega / \Omega_D)^{1/\alpha}$ with $\alpha \simeq 1/2$.
This yields $\Omega_\mathrm{max} \sim \Omega_D (p-1)^{\alpha}$, and the
shortest typical time scale $\tau$ of spontaneous intensity fluctuations can be estimated as
$\tau \sim \Omega_\mathrm{max}^{-1} \sim \Omega_D^{-1} (p - 1)^{-\alpha}$.
At high frequencies $\Omega \gg \Omega_D$ we find
$p \sim (\Omega / \Omega_D)^{1/2}$,
$\Omega_\mathrm{max} \sim \Omega_D p^2$, and $\tau \sim \Omega_D^{-1} p^{-2}$,
respectively. The latter results, on the contrary, are weakly sensitive to the peculiarities
of the sample, since for the fast oscillations the feedback mechanism is due
to relatively short diffusion paths that do not reach the boundaries of the sample.

The rise of the instability threshold with $\Omega$ can be qualitatively
understood by considering the phase difference $\Delta \phi(\Delta t)$
between two waves traveling through the disordered sample along the same diffusion path
but separated in
time by $\Delta t \sim \Omega^{-1}$. If $\varepsilon(\vec{r}, t)$
changes slowly with time, $\Delta \phi(\Delta t)$ comprises two contributions:
$\Delta \phi_L(\Delta t)$, which is the phase difference in a linear medium, and $\Delta \phi_{NL}(\Delta t)$,
which is the additional phase difference due to the nonlinearity. The second moment of the latter is
\begin{eqnarray}
\left< \Delta \phi_{NL}^2 (\Delta t) \right> &\sim& k_0^2 \varepsilon_2^2
\int_0^s \mathrm{d} s_1 \int_0^s \mathrm{d} s_2
\nonumber \\
&\times& \left< \Delta I(\vec{r}_1, \Delta t) \Delta I(\vec{r}_2, \Delta t) \right>,
\label{dphi}
\end{eqnarray}
where $s_i$ is a curvilinear coordinate of the point $\vec{r}_i$, the integrals are along the diffusion
path of typical length $s \sim L^2/l$, and
$\Delta I(\vec{r}_i, \Delta t)$ denotes the change of the intensity at $\vec{r}_i$ during the time $\Delta t$.
For $\Omega \ll \Omega_D$ we can assume that
$\left< \Delta I(\vec{r}_1, \Delta t) \Delta I(\vec{r}_2, \Delta t) \right> \sim
\left< \Delta \phi_{L}^2 (\Delta t) \right> \left< \delta I(\vec{r}_1, 0)
\delta I(\vec{r}_2, 0) \right>$ \cite{spivak00}.
Taking
$\left< \Delta \phi_{NL}^2 (\Delta t) \right> \agt \left< \Delta \phi_{L}^2 (\Delta t) \right>$ to be
the instability condition for the excitation of the speckle pattern at frequency $\Omega$,
and noting that
$\left< \delta I(\vec{r}_1, 0) \delta I(\vec{r}_2, 0) \right>
\sim I_0^2 / (k_0^2 l \left| \vec{r}_1 - \vec{r}_2 \right|)$ \cite{zyuzin87},
we recover $p \agt 1$
as the instability criterion. If $\Omega \gg \Omega_D$,
the long-range intensity correlation
establishes only for $\left| \vec{r}_1 - \vec{r}_2 \right| \alt (D \Delta t)^{1/2}$, and
the instability condition becomes $p \agt (\Omega / \Omega_D)^{1/2} \gg 1$.

The positive sign of the maximal Lyapunov exponent $\Lambda_{\mathrm{max}}$
for $p > 1$ (solid line in the inset of Fig.\ \ref{fig3}), as well as the
continuous spectrum
$0 < \Omega < \Omega_{\mathrm{max}}$ of frequencies with
$\Lambda > 0$, are hallmarks of chaotic behavior \cite{chaos}.
A sharp transition of the speckle pattern to chaos at $p = 1$
is reminiscent of the behavior observed in nonlinear systems with large (infinite) number of degrees of
freedom (e.g., random neural networks with an infinitely large number of nodes \cite{somp88}) and
should be contrasted to the ``route to chaos'' through a sequence of bifurcations, characteristic of
low-dimensional nonlinear systems \cite{chaos}.
As one can see from Fig.\ \ref{fig3}, the scaling of $\Lambda_\mathrm{max}$
with $p - 1$ appears to be roughly linear for  $\left| p - 1 \right| \ll 1$:
$\Lambda_\mathrm{max} \sim \Omega_D ( p - 1 )^{\beta}$ with
$\beta \simeq 1$.

To demonstrate the sensitivity of the results obtained for $\Omega \ll \Omega_D$
to the peculiarities of the disordered sample, we briefly
consider a sample with reflecting boundaries (dashed lines in Fig.\ \ref{fig3}).
All calculations can be carried out in the same way as for the sample with
open boundaries (see Appendix \ref{appa}), assuming that the Green's function
of Eq.\ (\ref{langevin}) is approximately the same as in the infinite medium:
$G(\vec{r}, \vec{r}_1; \Delta t) \simeq G_0(\vec{r}, \vec{r}_1; \Delta t)$.
We find that the absolute instability threshold $p_\mathrm{c}$
is roughly 2 times lower than in the open geometry,
$\alpha \simeq \beta \simeq 2$, and $\Lambda_\mathrm{max} = 0$ for $p < p_\mathrm{c}$.
For an arbitrary sample of disordered nonlinear medium
we expect $\Omega_\mathrm{max} \sim \Omega_D (p - p_\mathrm{c})^{\alpha}$ and
$\Lambda_\mathrm{max} \sim \Omega_D ( p - p_\mathrm{c} )^{\beta}$ for
$p - p_\mathrm{c} \ll 1$ and $p > p_\mathrm{c}$,
where $p_c \simeq 1$, $1/2 \alt \alpha \alt 2$, and
$1 \alt \beta \alt 2$.
By analogy \cite{chaos} with the theory of phase transitions,
$\Lambda_\mathrm{max}$ and $\beta$ can be identified with
the order parameter and the critical exponent, respectively.

Finally, it is worthwhile to note that
Eqs.\ (\ref{langevin}) and (\ref{djnl}) can also be derived from a
time-dependent disordered nonlinear
Schr\"{o}dinger equation with
a potential $u(\vec{r}) + g \left| \psi(\vec{r}, t) \right|^2$.
Upon the substitutions $\omega_0 \rightarrow E/\hbar$,
$k_0^2 \rightarrow 2 m E/\hbar^2$, $\varepsilon(\vec{r}) \rightarrow [-u(\vec{r})/E]$,
and $\varepsilon_2 \rightarrow (-g/E)$
[where
$E$ is the energy of the incident Schr\"{o}dinger wave, $m$ is the particle mass,
$u(\vec{r})$ is the disordered potential, and $g$ is the nonlinear constant],
our analysis is therefore valid in this case too.
The analogy between the wave equation (\ref{weq}) and the
Schr\"{o}dinger equation is known for the stationary case, when the solution
$\psi(\vec{r}, t)$ can be represented as $\psi_0(\vec{r}) \exp(-i \omega_0 t)$.
However, the dynamic solutions of the two equations differ due to different
dispersion relations.
Although the present paper deals with dynamic speckle patterns, their temporal
fluctuations are assumed to be slow and the analogy between the wave and
Schr\"{o}dinger equations is recovered within the accuracy of our analysis.

\begin{acknowledgments}
The author is grateful to R. Maynard and B.A. van Tiggelen for fruitful discussions and careful reading
of the manuscript. A.Yu. Zyuzin is acknowledged for a communication explaining
some details of Ref.\ \onlinecite{spivak00}.
\end{acknowledgments}

\appendix
\section{Derivation of Eq.\ (\ref{final})}
\label{appa}

In this Appendix we provide a derivation of Eq.\ (\ref{final})
from Eqs.\ (\ref{langevin}) and (\ref{djnl}).
Substituting $\delta I (\vec{r}, t) =  \delta I (\vec{r}, \nu) \exp( i \nu t )$
and $\vec{j}_\mathrm{ext} (\vec{r}, t)
= \vec{j}_\mathrm{ext} (\vec{r}, \nu) \exp( i \nu t )$ into
the two latter equations, we obtain
\begin{eqnarray}
&&i \nu \delta I(\vec{r}, \nu) - D \nabla^2 \delta I(\vec{r}, \nu) =
- \vec{\nabla} \cdot {\vec j}_\mathrm{ext}(\vec{r}, \nu),
\label{dinu}
\\
&&i \nu \vec{j}_\mathrm{ext} (\vec{r}, \nu) = i \nu \varepsilon_2
\int_V \mathrm{d}^3 \vec{r}^{\, \prime} \int_{0}^{\infty} \mathrm{d} \Delta t \,
\vec{q}( \vec{r}, \vec{r}^{\, \prime}, \Delta t) \;
\nonumber \\
&&\times \delta I(\vec{r}^{\, \prime}, \nu) \exp(-i \nu \Delta t).
\label{jnu}
\end{eqnarray}
If $\nu = 0$, Eq.\ (\ref{jnu}) is trivial and the statistical properties
of $\vec{j}_\mathrm{ext} (\vec{r}, 0)$ are determined by Eq.\ (\ref{current})
with $t = t_1$,
the same equation as in the case of the linear medium, while the static, time-independent
part of the intensity fluctuation $\delta I(\vec{r}, 0)$ is found by solving the
stationary Langevin equation [Eq.\ (\ref{dinu}) with $\nu = 0$]. Hence,
the time-independent part of the speckle pattern remains the same as in the
linear medium. If, in contrast, $\nu \ne 0$ (as we assume in the main text),
we divide both sides of Eq.\ (\ref{jnu}) by $i \nu$, multiply the $i$-th
Cartesian component of the resulting
equation by the $j$-th Cartesian component of a similar equation
for $\vec{j}_\mathrm{ext}^* (\vec{r}_1, \nu)$,
and average the result over disorder. This yields
\begin{eqnarray}
&&\left< j_\mathrm{ext}^{(i)} (\vec{r}, \nu)
j_\mathrm{ext}^{(j) *} (\vec{r}_1, \nu) \right>
\nonumber \\
&&=\varepsilon_2^2
\int_V \mathrm{d}^3 \vec{r}^{\, \prime}
\int_V \mathrm{d}^3 \vec{r}_1^{\, \prime}
\left< \delta I (\vec{r}^{\, \prime}, \nu) \delta I^* (\vec{r}_1^{\, \prime}, \nu)
\right>
\nonumber \\
&&\times\int_{0}^{\infty} \mathrm{d} \Delta t
\int_{0}^{\infty} \mathrm{d} \Delta t_1 \,
\left< q^{(i)}( \vec{r}, \vec{r}^{\, \prime}, \Delta t)
q^{(j)*}( \vec{r}_1, \vec{r}_1^{\, \prime}, \Delta t_1) \right>
\nonumber \\
&&\times\exp[-i \nu \Delta t + i \nu^* \Delta t_1)],
\label{jjnu}
\end{eqnarray}
where $j_\mathrm{ext}^{(i)} (\vec{r}, \nu)$ denotes the
$i$-th Cartesian component of $\vec{j}_\mathrm{ext} (\vec{r}, \nu)$.
After the substitution of Eq.\ (\ref{qcorr}) for
$\left< q^{(i)}( \vec{r}, \vec{r}^{\, \prime}, \Delta t)
q^{(j)*}( \vec{r}_1, \vec{r}_1^{\, \prime}, \Delta t_1) \right>$,
the time integrations in Eq.\ (\ref{jjnu}) yield
\begin{eqnarray}
&&\int_{0}^{\infty} \mathrm{d} \Delta t
\int_{0}^{\infty} \mathrm{d} \Delta t_1 \,
\left< q^{(i)}( \vec{r}, \vec{r}^{\, \prime}, \Delta t)
q^{(j)*}( \vec{r}_1, \vec{r}_1^{\, \prime}, \Delta t_1) \right>
\nonumber \\
&&\times \exp[-i \nu \Delta t + i \nu^* \Delta t_1)]
\nonumber \\
&&= 3 \pi D^2 (c^2/l) \delta_{i j} \delta(\vec{r} - \vec{r}_1)
\nonumber \\
&&\times
\left[ \left< I(\vec{r}^{\, \prime}) \right>
G(\vec{r}^{\, \prime}, \vec{r}_1^{\, \prime}; \nu)
G^*(\vec{r}_1^{\, \prime}, \vec{r}; \nu - \nu^*) \left< I(\vec{r}) \right>
\right.
\nonumber \\
&&+ \left. \left< I(\vec{r}_1^{\, \prime}) \right>
G^*(\vec{r}_1^{\, \prime}, \vec{r}^{\, \prime}; \nu)
G(\vec{r}^{\, \prime}, \vec{r}; \nu - \nu^*) \left< I(\vec{r}) \right> \right.
\nonumber \\
&&- \left.
\left< I(\vec{r}^{\, \prime}) \right> G(\vec{r}^{\, \prime}, \vec{r}; \nu)
\left< I(\vec{r}_1^{\, \prime}) \right>
G^*(\vec{r}_1^{\, \prime}, \vec{r}; \nu) \right],
\label{timeint}
\end{eqnarray}
where $G(\vec{r}, \vec{r}_1; \nu)$ is the Fourier transform of
$G(\vec{r}, \vec{r}_1; \Delta t)$.
Equation (\ref{jjnu}) can now be rewritten as
\begin{eqnarray}
\left< j_\mathrm{ext}^{(i)} (\vec{r}, \nu)
j_\mathrm{ext}^{(j) *} (\vec{r}_1, \nu) \right> &=&
A(\vec{r}, \nu) \delta_{ij} \delta(\vec{r}-\vec{r}_1),
\label{jjnu2}
\end{eqnarray}
where
\begin{eqnarray}
&&A(\vec{r}, \nu) =
3 \pi D^2 (c^2/l) \varepsilon_2^2
\int_V \mathrm{d}^3 \vec{r}^{\, \prime}
\int_V \mathrm{d}^3 \vec{r}_1^{\, \prime}
\nonumber \\
&&\times
\left< \delta I (\vec{r}^{\, \prime}, \nu) \delta I^* (\vec{r}_1^{\, \prime}, \nu)
\right>
\nonumber \\
&&\times \left[ \left< I(\vec{r}^{\, \prime}) \right>
G(\vec{r}^{\, \prime}, \vec{r}_1^{\, \prime}; \nu)
G^*(\vec{r}_1^{\, \prime}, \vec{r}; \nu - \nu^*) \left< I(\vec{r}) \right>
\right.
\nonumber \\
&&+ \left. \left< I(\vec{r}_1^{\, \prime}) \right>
G^*(\vec{r}_1^{\, \prime}, \vec{r}^{\, \prime}; \nu)
G(\vec{r}^{\, \prime}, \vec{r}; \nu - \nu^*) \left< I(\vec{r}) \right> \right.
\nonumber \\
&&- \left.
\left< I(\vec{r}^{\, \prime}) \right> G(\vec{r}^{\, \prime}, \vec{r}; \nu)
\left< I(\vec{r}_1^{\, \prime}) \right>
G^*(\vec{r}_1^{\, \prime}, \vec{r}; \nu) \right].
\label{a}
\end{eqnarray}
In the following, we replace both $\left< I(\vec{r}) \right>$ and
$A(\vec{r}, \nu)$ by their spatial averages $I_0$ and $A(\nu)$,
respectively.
This simplifies the further analysis considerably, while can only affect the
final result by a numerical factor of order unity, since
$\left< I(\vec{r}) \right>$ and
$A(\vec{r}, \nu)$ do not change significantly as long as the point $\vec{r}$
is far enough from the sample boundaries.

We now admit that Eq.\ (\ref{jjnu2}) for the correlation function of Langevin
currents at $\nu \ne 0$ in a nonlinear medium has a form similar
to Eq.\ (\ref{current}) for Langevin
currents in a linear medium.
This allows us to proceed with analysis of Eq.\ (\ref{dinu})
in the same way as it was done for Eq.\ (\ref{langevin}) in the linear medium
\cite{zyuzin87}.
To simplify further calculations, we assume that the disordered sample has
open boundaries (i.e. that the multiple-scattered waves leave the sample when they
reach the boundary) and hence the Green's function
of Eq.\ (\ref{langevin}), $G(\vec{r}, \vec{r}_1; \Delta t)$,
can be approximately written as
$G_0(\vec{r}, \vec{r}_1; \Delta t) \exp(-\Omega_D \Delta t)$, where
$G_0(\vec{r}, \vec{r}_1; \Delta t) = (4 \pi D \Delta t)^{-3/2}
\exp[-\left| \vec{r} - \vec{r}_1  \right|^2/(4 D \Delta t)]$
is the Green's function in the infinite medium, $\exp(-\Omega_D \Delta t)$
describes the leakage of the wave through the sample boundaries, and
$\Omega_D = D/L^2$.
We now write the solution of Eq.\ (\ref{dinu}) as
\begin{eqnarray}
\delta I(\vec{r}, \nu) &=&
-\int_V \mathrm{d}^3 \vec{r}^{\, \prime} G(\vec{r}, \vec{r}^{\, \prime}; \nu)
\left[ \vec{\nabla} \cdot \vec{j}_{\mathrm{ext}}(\vec{r}^{\, \prime}, \nu) \right]
\nonumber \\
&=& \int_V \mathrm{d}^3 \vec{r}^{\, \prime}
\left[ \vec{\nabla} G(\vec{r}, \vec{r}^{\, \prime}; \nu)
\cdot \vec{j}_{\mathrm{ext}}(\vec{r}^{\, \prime}, \nu) \right],
\label{di}
\end{eqnarray}
where the second line is obtained as a result of integration by parts, assuming
$G(\vec{r}, \vec{r}^{\, \prime}; \nu)
j^{(i)}_{\mathrm{ext}}(\vec{r}^{\, \prime}, \nu) = 0$ at the boundary of the
disordered sample. Multiplying Eq.\ (\ref{di}) by a similar equation for
$\delta I^*(\vec{r}_1, \nu)$, performing the averaging over disorder using
Eq.\ (\ref{jjnu2}), and carrying out the necessary integrations,
we obtain
\begin{eqnarray}
&&\left< \delta I(\vec{r}, \nu) \delta I^*(\vec{r}_1, \nu) \right>
\nonumber \\
&&=
\frac{A(\nu)}{D} \left[ \mathrm{Re} G(\vec{r}, \vec{r}_1; \nu) -
\frac{\mathrm{Im} \nu}{\mathrm{Re} \nu} \mathrm{Im} G(\vec{r}, \vec{r}_1; \nu)
\right].
\label{didi}
\end{eqnarray}

Substituting Eq.\ (\ref{didi}) into Eq.\ (\ref{a}), dividing both sides of the
resulting equation by $A(\nu) \ne 0$, recalling that
$\nu = \Omega - i \Lambda$, and performing a change of variables
$\vec{R} = [\Omega/(2 D)]^{1/2} \vec{r}$ (and similarly for
$\vec{r}^{\, \prime}$ and $\vec{r}_1^{\, \prime}$), we obtain
\begin{eqnarray}
1 = C_1 \Delta n^2 (L/l)^3
h(\vec{R}, \Omega/\Omega_D, \Lambda/\Omega_D),
\label{eq1}
\end{eqnarray}
where $C_1$ is a numerical constant,
$\Delta n = (\varepsilon_2/2) I_0$, and
the dimensionless function $h$ is defined as
\begin{eqnarray}
&&h(\vec{R}, \Omega/\Omega_D, \Lambda/\Omega_D)
= \left( \Omega_D/\Omega \right)^{3/2}
\int \mathrm{d}^3 \vec{R}^{\, \prime}
\int \mathrm{d}^3 \vec{R}_1^{\, \prime}
\nonumber \\
&&\times \left[
{\cal G}(\vec{R}^{\, \prime}, \vec{R}_1^{\, \prime}; \gamma)
{\cal G}_1^*(\vec{R}_1^{\, \prime}, \vec{R}; \gamma_1)
\right. \nonumber \\
&&+ \left. {\cal G}^*(\vec{R}_1^{\, \prime}, \vec{R}^{\, \prime}; \gamma)
{\cal G}_1(\vec{R}^{\, \prime}, \vec{R}; \gamma_1) \right.
\nonumber \\
&&- \left.
{\cal G}(\vec{R}^{\, \prime}, \vec{R}; \gamma)
{\cal G}^*(\vec{R}_1^{\, \prime}, \vec{R}; \gamma) \right]
\nonumber \\
&&\times \left[ \mathrm{Re}{\cal G}(\vec{R}^{\, \prime}, \vec{R}_1^{\, \prime}; \gamma)
+ \frac{\Lambda}{\Omega}
\mathrm{Im}{\cal G}(\vec{R}^{\, \prime}, \vec{R}_1^{\, \prime}; \gamma)
\right],
\label{eq2}
\end{eqnarray}
\begin{eqnarray}
&&{\cal G}(\vec{R}, \vec{R}_1; \gamma)
\nonumber \\
&&= \frac{1}{\left| \vec{R} - \vec{R}_1 \right|}
\exp\left[-\left(\gamma + i/\gamma \right) \left| \vec{R} - \vec{R}_1 \right|
\right],
\label{g}
\\
&&{\cal G}_1(\vec{R}, \vec{R}_1; \gamma_1)
\nonumber \\
&&=
\frac{1}{\left| \vec{R} - \vec{R}_1 \right|}
\exp\left(-2 \gamma_1 \left| \vec{R} - \vec{R}_1 \right| \right),
\label{g1}
\\
&&\gamma = \left\{ \left[ 1 + \left( \frac{\Lambda + \Omega_D}{\Omega} \right)^2
\right]^{1/2} \right.
\nonumber \\
&&+ \left. \frac{\Lambda + \Omega_D}{\Omega} \right\}^{1/2},
\label{gamma}
\\
&&\gamma_1 = \left( \frac{\Lambda + \Omega_D/2}{\Omega} \right)^{1/2}.
\label{gamma1}
\end{eqnarray}

We now assume that the disordered sample has the shape of a sphere centered at the
origin and that $h(\vec{0}, \Omega/\Omega_D, \Lambda/\Omega_D)$ provides a good
estimation of $h$ for the points $\vec{R}$ located far enough from the
boundaries. Defining
$F(\Omega/\Omega_D, \Lambda/\Omega_D) =
h(\vec{0}, 0, 0) / h(\vec{0}, \Omega/\Omega_D, \Lambda/\Omega_D)$
and
introducing the effective nonlinearity parameter
$p = \Delta n^2 (L/l)^3$, we rewrite Eq.\ (\ref{eq1}) as
\begin{eqnarray}
p = C_2 F(\Omega/\Omega_D, \Lambda/\Omega_D),
\label{eq3}
\end{eqnarray}
where $C_2$ is a numerical factor of order unity.
Since we have already made some approximations that affect the final result by
a numerical factor of order unity (e.g., we replaced $\left< I(\vec{r}) \right>$ by
$I_0$), we omit $C_2$ in Eq.\ (\ref{eq3}) and obtain Eq.\ (\ref{final})
of the main text.
The most of integrations in Eq.\ (\ref{eq2}) can be performed
analytically, while the remaining integrations are easily carried out numerically,
allowing us to determine the value of $\Lambda$ for given $p$, $\Omega_D$, and
$\Omega$ from Eq.\ (\ref{eq3}).

An important comment is in order in connection with Eq.\ (\ref{eq3}) and the analysis
it results from. The correlation function of intensity fluctuations
$\left< \delta I (\vec{r}, \nu) \delta I^* (\vec{r}_1, \nu)
\right>$ entering into Eqs.\ (\ref{jjnu}) and (\ref{a}) contains,
in principle, not only the long-range
contribution given by Eq.\ (\ref{didi}), but also a short-range one
$\left< \delta I (\vec{r}, \nu) \delta I^* (\vec{r}_1, \nu)
\right> \sim (l/k_0^2) I_0^2 \delta(\vec{r} - \vec{r}_1)$.
The latter contribution has been
neglected in our analysis, which is justified for large enough sample size
($L/l \gg k_0 l$) and moderate frequency
$\Omega \ll \Omega_D [L / (k_0 l^2)]^2$.
If one of the above inequalities is violated, the roles played by the
short- and long-range contributions to the correlation function of intensity
fluctuations in development of the instability become comparable,
and the above analysis is no longer valid.



\begin{thebibliography}{99}

\bibitem{rossum99}
M.C.W. ~van~Rossum and Th.M.~Nieuwenhuizen,
\textit{Rev. Mod. Phys.} \textbf{71,} 313 (1999).

\bibitem{berk94}
R.~Berkovits and S.~Feng, \textit{Phys. Rep.} \textbf{238,} 135 (1994).

\bibitem{sebbah01}
\textit{Waves and Imaging through Complex Media,}
edited by P.~Sebbah (Kluwer, Dordrecht, 2001).

\bibitem{agran91}
V.M.~Agranovich and V.E.~Kravtsov,
\textit{Phys. Rev. B} \textbf{43,} 13691 (1991);
A.~Heidereich, R.~Maynard, and B.A.~van~Tiggelen,
\textit{Opt. Commun.} \textbf{115,} 392 (1995);
R.~Bressoux and R.~Maynard, in Ref.\ \onlinecite{sebbah01}, p.~445.

\bibitem{kravtsov91}
V.E.~Kravtsov, V.M.~Agranovich, and K.I.~Grigorishin,
\textit{Phys. Rev. B} \textbf{44,} 4931 (1991).

\bibitem{kravtsov90}
V.E.~Kravtsov, V.I.~Yudson, and V.M.~Agranovich,
\textit{Phys. Rev. B} \textbf{41,} 2794 (1990);
J.C.J.~Paasschens \textit{et al.,}
\textit{Phys. Rev. A} \textbf{56,} 4216 (1997).

\bibitem{boer93}
J.F.~de~Boer \textit{et al.,}
\textit{Phys. Rev. Lett.} \textbf{71,} 3947 (1993).

\bibitem{bress00}
R.~Bressoux and R.~Maynard,
\textit{Europhys. Lett.} \textbf{50,} 460 (2000).

\bibitem{wonderen94}
A.J.~van~Wonderen,
\textit{Phys. Rev. B} \textbf{50,} 2921 (1994).

\bibitem{tomita01}
M.~Tomita, T.~Ito, and S.~Hattori,
\textit{Phys. Rev. B} \textbf{64,} 180202 (2001).

\bibitem{spivak00} B.~Spivak and A.~Zyuzin,
\textit{Phys. Rev. Lett.} \textbf{84,} 1970 (2000).

\bibitem{skip00} S.E.~Skipetrov and R.~Maynard,
\textit{Phys. Rev. Lett.} \textbf{85,} 736 (2000);
S.E.~Skipetrov,
\textit{Phys. Rev. E} \textbf{63,} 056614 (2001).

\bibitem{shen84}
Y.R.~Shen, \textit{The Principles of Nonlinear Optics}
(Wiley, New York, 1984);
G.S.~He and S.H.~Liu,
\textit{Physics of Nonlinear Optics}
(World Scientific, Singapore, 1999).

\bibitem{ishim81}
A.~Ishimaru,
\textit{Wave Propagation and Scattering in Random Media}
(Academic Press, New York, 1978).

\bibitem{zyuzin87}
A.Yu.~Zyuzin and B.Z.~Spivak, \textit{Sov. Phys. JETP} \textbf{66,} 560 (1987);
R.~Pnini and B.~Shapiro,
\textit{Phys. Rev. B} \textbf{39,} 6986 (1989).

\bibitem{gibbs85}
H.M.~Gibbs,
\textit{Optical Bistability: Controlling Light With Light}
(Academic, New York, 1985)

\bibitem{arecchi99}
F.T.~Arecchi, S.~Boccaletti, and P.~Ramazza,
\textit{Phys. Rep.} \textbf{318,} 1 (1999).

\bibitem{chaos}
See, e.g., H.~G.~Schuster, \textit{Deterministic Chaos} (Physik-Verlag, Weinheim, 1984) or
E.~Ott, \textit{Chaos in Dynamical Systems} (Cambridge University Press, Cambridge, 1993).

\bibitem{somp88}
H.~Sompolinsky, A.~Crisanti, and H.~J.~Sommers,
\textit{Phys. Rev. Lett.} \textbf{61,} 259 (1988).

\end{thebibliography}
\end{document}